# Modelling periodic structure formation on 100Cr6 steel after irradiation with femtosecond-pulsed laser beams


George D. Tsibidis [1♣], Alexandros Mimidis [1,2], Evangelos Skoulas [1,2], Sabrina V. Kirner [3], Jörg Krüger [3], Jörn Bonse [3♣] and Emmanuel Stratakis [1,2♣]

[1] *Institute of Electronic Structure and Laser (IESL), Foundation for Research and Technology (FORTH), N. Plastira 100, Vassilika Vouton, 70013, Heraklion, Crete, Greece*

[2] *Materials Science and Technology Department, University of Crete, 71003 Heraklion, Greece*

[3] *Bundesanstalt für Materialforschung und -prüfung (BAM), Unter den Eichen 87, 12205 Berlin, Germany*


## ABSTRACT


We investigate the periodic structure formation upon intense femtosecond pulsed irradiation of chrome steel (100Cr6) for linearly polarised laser beams. The underlying physical mechanism of the laser induced periodic structures is explored, their spatial frequency is calculated and theoretical results are compared with experimental observations. The proposed theoretical model comprises estimations of electron excitation, heat transfer, relaxation processes, and hydrodynamics-related mass transport. Simulations describe the sequential formation of sub-wavelength ripples and supra-wavelength grooves. In addition, the influence of the laser wavelength on the periodicity of the structures is discussed. The proposed theoretical investigation offers a systematic methodology towards laser processing of steel surfaces with important applications.


♣ tsibidis@iesl.forth.gr

♣ joern.bonse@bam.de

♣ stratak@iesl.forth.gr




## 1. Introduction

Material processing with ultra-short pulsed lasers has received considerable attention over the past decades due to its important technological applications, in particular in industry and medicine [1-8]. A laser-based technique allows an efficient methodology of material engineering as it enables precise tailoring of surface features with significant influence on optical properties, tribological performance and wettability [2, 3, 9-12].

One type of surface modification, the production of laser-induced periodic surface structures (LIPSS, ripples) on solids has been studied intensively for linearly polarized beams. LIPSS formation is a universal effect and they have been observed in all types of materials (metals, semiconductors, dielectrics, polymers). Previous theoretical approaches or experimental observations related to the underlying physical mechanisms of the formation of these structures were performed in sub-melting [13] or ablation conditions [14-19]. With respect to ripples, various mechanisms have been proposed to account for their formation: interference of the incident wave with an electromagnetic wave scattered at the rough surface [15, 17, 20], with a surface plasmon wave (SPW) [16, 19, 21-24], or due to self-organisation mechanisms [25]. The main focus of investigation centered on structures with periodicities comparable to the laser wavelength ($\approx \lambda_L$) which are termed *low spatial frequency LIPSS* [LSFL]. These structures have an orientation perpendicular or parallel to the polarisation of the incident beam depending on the material type ([20, 26-28] and references therein).

Nevertheless, another type of experimentally observed periodic structures characterized by an always parallel orientation to the polarisation of the laser and, substantially smaller spatial frequency has yet to be investigated in all types of materials. A physical mechanism that explains the formation of these periodic structures, the so-called *grooves*, was recently presented in semiconductors [29, 30] and dielectrics [20], which attributes their development predominantly to hydrodynamical rather than SPW related effects, as the laser-produced density of excited carriers is not sufficient to induce significant SPW. It is, therefore, questionable whether similar laser-induced supra-wavelength structures on metals are generated from a different mechanism, as electron densities in metals are assumed to be constant. It is of paramount importance, both from a



fundamental point of view and industrial applicability, to present a theoretical framework of the underlying process of the groove formation that also predicts the '*LIPSS to grooves*' transition.

One material with extremely important technological applications is steel. It is characterised by a high tensile strength, which is a significant feature in a wide range of applications such as automotive, construction, domestic appliances, and mechanical machinery. Although some previous work on morphological effects on steel after fs-laser irradiation has been conducted by many groups (just see [31] as an example for a systematic variation of laser processing parameters such as fluence and pulse number), sufficient knowledge of the underlying physical mechanisms that correlate laser-matter interaction with observable effects on steel is still missing. Therefore, a theoretical modelling of the laser interaction with steel can provide significant details about the thermal and mechanical response of the material. Herein, the physical mechanisms for the formation of a variety of periodic structures in one particular type of chrome steel, 100Cr6, under intense femtosecond laser conditions will be explored. A theoretical investigation is conducted under the assumption that the laser conditions are sufficient to induce mass removal (ablation) and melting of a portion of the material. The proposed theoretical framework assumes electron and SPW excitation upon laser irradiation, electron-phonon relaxation processes, phase change (melting), mass removal (ablation), and resolidification. Special emphasis is drawn on the formation mechanisms of LIPSS and grooves, while theoretical predictions will be tested against experimental observations.

## 2. Experimental protocol

Circular slabs of hardened 100Cr6 steel (24 mm diameter, 8 mm thickness) with a polished surface (roughness $R_a$ = 35 nm, $R_{rms}$ = 48 nm) were purchased from Optimol Instruments Prüftechnik GmbH (München, Germany).

At FORTH the samples were irradiated with a beam of 1026 nm (1 kHz repetition rate) and 513 nm (60 kHz repetition rate) central wavelengths with a pulse duration of



170 fs. All samples were ultrasonically cleaned using isopropanol for 10 minutes both prior to and following irradiation. Then samples were investigated under an electron microscope (JEOL JSM7500F) while LIPSS periodicities were measured by two dimensional fast Fourier transform (2D-FFT) analysis on the acquired scanning electron microscopy (SEM) micrographs using the Gwyddion software.

At BAM, the samples were mounted on a motorized $x$-$y$-$z$-linear translation stage and placed normal to the incident laser beam close to the focal position of a spherical dielectric mirror with a focal length of 500 mm. At this sample position, a Gaussian beam waist $1/e^2$-radius of $w_0 = 65$ µm was determined using a method proposed by Liu [32] for the focused laser beam ($\tau_p = 30$ fs pulse duration). Spot processing was performed at a fixed number of pulses ($NP$) and a chosen peak fluence $F$ (in front of the sample), calculated from the laser pulse energy $E$ via $F = 2E/(\pi w_0^2)$ [32]. The laser radiation was emitted from a Ti:sapphire amplifier system (Compact Pro, Femtolasers, Vienna, Austria) operated at a center wavelength of $\lambda_L = 800$ nm and at $f = 1$ kHz pulse repetition rate.

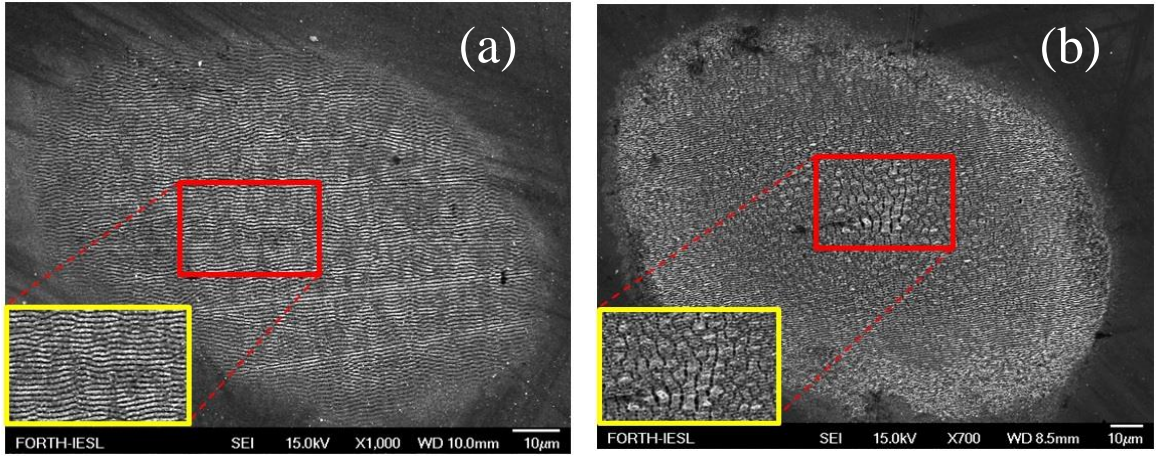

FIG. 1 (Color Online). SEM images showing ripples and /or grooves on 100Cr6 for $\lambda_L = 1026$ nm (a) $NP = 20$, $F = 0.22$ J/cm² and (b) $NP = 150$, $F = 0.72$ J/cm². Insets (in *yellow*) show the surface modification in the spot center. The polarization direction is vertical here.



After laser processing, the samples were cleaned in acetone for three minutes using an ultrasonic bath and characterized by SEM (Carl Zeiss Gemini Supra 40). The most frequent periods and the period ranges of the fs-laser processed ripples and grooves were determined from the corresponding micrographs by either one- and two-dimensional Fourier transforms (ripples) or averaged over six manual period measurements (grooves). While the Fourier transforms offer information contained in a central area of $50 \times 50\ \mu m^2$ of the corresponding SEM images, the manual groove period measurements were performed at six individual grooves at the center of the SEM image, providing rather an estimation.

Experimental results are illustrated in Figs. 1 and 2, where the SEM images show formation of ripples and/or grooves that were produced on the surface of the irradiated zone. It is evident that the grooves have a larger periodicity than the ripples and are oriented perpendicular to the sub-wavelength structures while fine rippled structures are produced after irradiation with a small number of pulses. To provide a systematic correlation of the laser beam conditions with the observed surface structures, it is important to follow a detailed analysis of the underlying physical mechanisms.

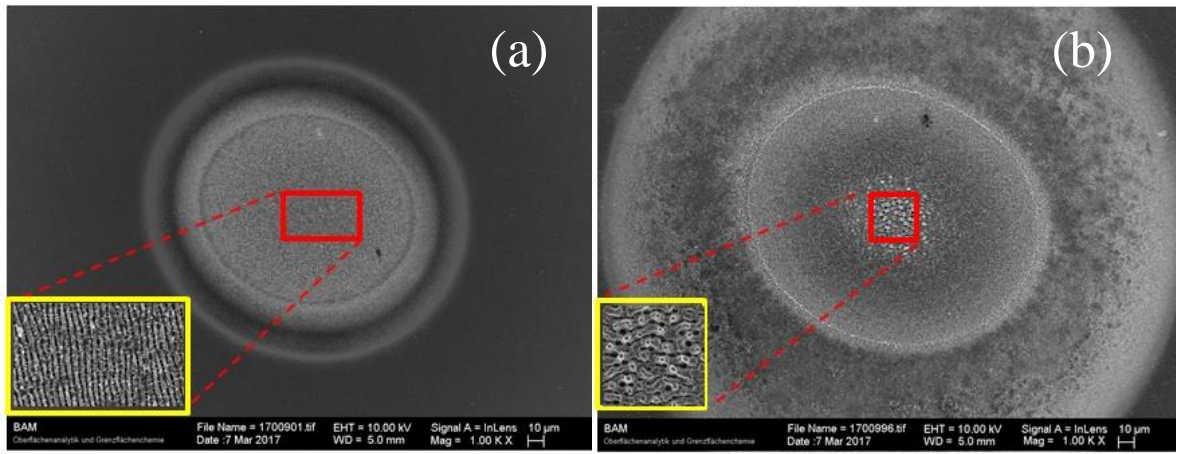

FIG. 2 SEM images showing ripples and /or grooves on 100Cr6 steel for $\lambda_L$ = 800 nm (a) $NP$ = 100, $F$ = 0.5 J/cm$^2$ and (b) $NP$ = 100, $F$ = 2.5 J/cm$^2$. Insets (in *yellow*) show the surface modification in the spot center. The polarization direction is horizontal here.



## 3. Theory

To understand the physical mechanism that accounts for the surface modification upon irradiation of metals with femtosecond (fs) pulsed lasers, we perform a multiscale modelling of the processes that describe laser beam energy absorption and response of the material. Therefore, the theoretical model that is presented should comprise the following components: (i) a term that describes energy absorption, (ii) a term that describes electron excitation, (iii) a heat transfer component that accounts for electron-lattice thermalisation through particle dynamics and heat conduction and carrier-phonon coupling, and (iv) a hydrodynamics component that describes fluid dynamics followed by a mass removal and re-solidification process in areas where a phase transition occurs. In principle, the processes start after some fs, they continue to mechanisms that complete after some picoseconds (ps) while others require more time and they last up to the nanosecond (ns) regime.

## 3.1 Energy absorption, electron excitation and relaxation processes.

The two-temperature model (TTM) constitutes the standard theoretical method to investigate laser-matter interaction upon femtosecond laser irradiation, which assumes an instantaneous electron excitation during the laser pulse that produces fast electron thermalisation on the femtosecond timescale [33]. The TTM is implemented by the following set of coupled differential equations that describe the absorption of optical radiation by the electrons and the energy transfer between the electron and lattice subsystems.

$$
\begin{aligned}
C_e \frac{\partial T_e}{\partial t} &= \vec{\nabla} \bullet (\kappa_e \vec{\nabla} T_e) - g_{e-ph}(T_e - T_L) + S(\vec{r}, \text{t}) \\
C_L \frac{\partial T_L}{\partial t} &= \vec{\nabla} \bullet (\kappa_L \vec{\nabla} T_L) + g_{e-ph}(T_e - T_L)
\end{aligned}
\tag{1}
$$

$T_e$ and $T_L$ are the electron and lattice temperatures, $\kappa_e$ and $\kappa_L$ (~0.01$\kappa_e$ and therefore, that term could even be neglected) stand for the electron and lattice heat conductivities, $C_e$ and $C_L$ correspond to the electron and lattice heat capacities, and $g_{e\text{-ph}}$ is the electron-phonon coupling strength. The heat source $S$ in a Cartesian system of coordinates $(x,y,z)$ is modelled by assuming a Gaussian temporal profile and it has the following (simplified, to first approximation) form [35]



$$S(x, y, z, t) = \frac{\alpha(1-R)\sqrt{4\log 2}\,F}{\sqrt{\pi}\,\tau_p} \exp\left(-4\log 2\left(\frac{t-3\tau_p}{\tau_p}\right)^2\right) \exp(-\alpha z) \exp\left[-\left(\frac{x+y}{w_0}\right)^2\right], \qquad (2)$$

where $\tau_p$ is the pulse duration, $F$ is the peak fluence, $\alpha$ and $R$ stand for the absorption coefficient and the reflectivity, respectively, of the irradiated material and $w_0$ is the $1/e$-Gaussian spot radius. For the sake of simplicity, it is assumed that optical properties ($\alpha$, $R$) do not have a spatio-temporal variation.

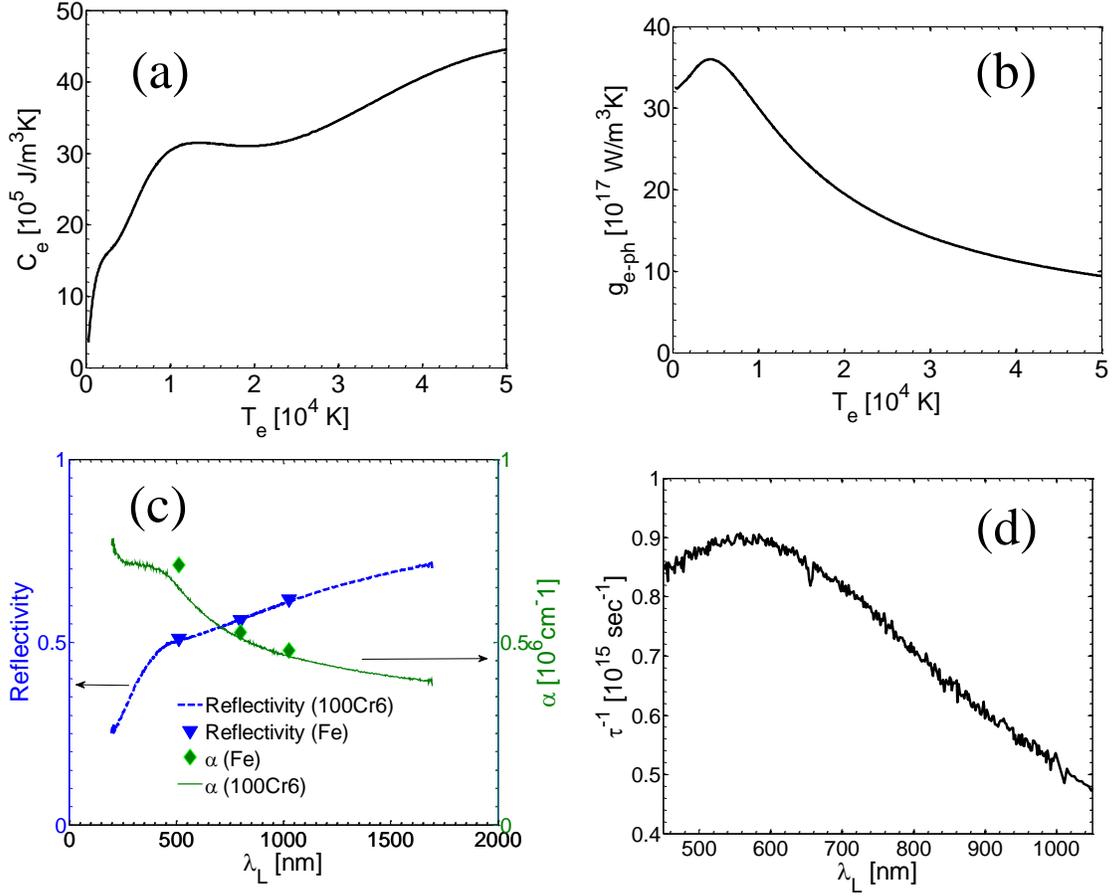

FIG. 3 (Color Online) Heat capacity $C_e$ (a) and electron-phonon coupling strength $g_{e\text{-ph}}$ (b) as a function of the electron temperature, for Fe, (c) absorption coefficient $\alpha$ and reflectivity $R$ (at normal incidence), and (d) reciprocal of electron relaxation time $\tau$, for 100Cr6 steel as a function of the laser wavelength $\lambda_L$ [34]. The lines in (c) are obtained from ellipsometric measurements of 100Cr6 steel, while the points represent data for pure iron taken from Ref. [39].

Equations (1-2) have been extensively used to describe the heat transfer from the electron system to the lattice system in various types of metals, semiconductors or dielectrics. Previous works have also provided a detailed calculation of the thermophysical properties of the material (i.e. electron heat capacity, conductivity, electron-phonon coupling constant) based on the computation of the density of states (DOS) for various energies below and above the Fermi energy [34]. Nevertheless, there is a lack of knowledge of these parameters for materials used for industrial applications such as 100Cr6 steel. A rigorous approach would be to use first principles and derive, firstly, the DOS for this material by using relevant software, density functional theory and experimental data [36] and, secondly, produce an estimate for those parameters. Herein, a simplified approach is followed in which an approximation is performed based on the fact that iron (Fe) is the main ingredient of the chrome steel [37]. Although, the precise estimation of the heat capacity and the electron-phonon coupling constant could significantly influence the maximum attained value of the electron temperature as well as the electron-phonon relaxation time and maximum lattice temperature, herein, it is assumed that the employment of the thermophysical properties based on the fitting of data for Fe does not differ significantly for 100Cr6. Indeed, recent results indicate that the temperature dependent electron heat capacity of a steel alloy is not substantially different from that predicted for Fe [38]. Similarly, our calculations indicate that a more rigorous computation of the electron-phonon coupling is not anticipated to produce substantially different morphological results.

Therefore, the (electron) temperature dependent heat capacity $C_e$ (Fig. 3a) and electron-phonon coupling strength $g_{e-ph}$ of Fe (Fig. 3b) are computed using a polynomial fitting of calculated values [34]. The heat conductivity is calculated assuming the



reciprocal of the electron relaxation time $\tau$ is the sum of the electron-electron and electron-phonon collision rates, $A(T_e)^2$ and $BT_L$, respectively [39]. Hence, the heat conductivity is computed by the expression $k_e = k_{eo} \dfrac{BT_e}{A(T_e)^2 + BT_L}$ [40]. The parameters $A$ and $B$, can be obtained from variable angle spectral ellipsometric measurements of the refractive index ($n$) and the extinction coefficient ($k$) of the polished 100Cr6 steel at various wavelengths. From that data, the corresponding wavelength dependent linear absorption coefficient $\alpha$ and the room temperature surface reflectivity $R$ were calculated (Fig. 3c) via the expressions $R=[(n-1)^2+k^2]/[(n+1)^2+k^2]$ and $\alpha=4\pi k/\lambda$. It is evident that the values for the optical properties of 100Cr6 steel for $\lambda_L = 513$ nm, 800 nm, and 1026 nm (the laser wavelength values used in this work) are very close to the ones reported for Fe [41]. This is an indication that the magnitude of the laser energy absorption and optical penetration depth dictated by $R$ and $\alpha$, respectively, for 100Cr6 are similar to those of Fe.

The reciprocal of electron relaxation time $\gamma$ ($\equiv \tau^{-1}$) is computed by the following expressions that relate the dielectric permittivity of the material ($\varepsilon$) to $n$ and $k$

$$\begin{aligned}
\varepsilon &= \varepsilon_r + i\varepsilon_i \\
\varepsilon_r &= n^2 - k^2 \\
\varepsilon_i &= 2nk
\end{aligned} \tag{3}$$

and [42]

$$\begin{aligned}
\varepsilon_r &= 1 - \frac{\omega_p^2}{\omega^2 + \gamma^2} \\
\varepsilon_i &= \frac{\omega_p^2 \gamma}{\left(\omega^2 + \gamma^2\right)\omega}
\end{aligned}, \tag{4}$$

where $\omega_p$ is the plasma frequency [43]. Estimates of the values of $A$ and $B$ are $0.98 \times 10^7$ $K^{-2}s^{-1}$ and $2.8 \times 10^{11}$ $K^{-1}s^{-1}$, respectively, based on a least-squares fitting procedure performed on the data shown in Fig. 3d.



## 3.2. Fluid dynamics and mass removal

In order to describe the thermal response of the material after electron-phonon relaxation processes, it is important to model heat transfer into the material, possible phase transition and solidification. Since the laser irradiation conditions used in the simulations are sufficient to melt a portion of the material (once the lattice temperature reaches the melting temperature $T_{melt}$), the induced phase change has to be thoroughly analysed. Therefore, the transport of the molten material is described by the assumption that fluid behaves as an incompressible Newtonian fluid and its dynamics is provided by the following set of equations (to account for the mass, energy, and momentum conservation, respectively) [19]

$$\vec{\nabla} \bullet \vec{u} = 0$$
$$\left(C_L^{(m)} + L_m \delta(T_L - T_{melt})\right) \frac{\partial T_L}{\partial t} + C_L^{(m)} \vec{\nabla} \bullet (\vec{u} T_L) = \vec{\nabla} \bullet (\kappa_L \vec{\nabla} T_L) \quad , \qquad (5)$$
$$\rho_L^{(m)} \left(\frac{\partial \vec{u}}{\partial t} + \vec{u} \bullet \vec{\nabla} \vec{u}\right) = \vec{\nabla} \bullet \left(-P + \mu^{(m)} (\vec{\nabla} \vec{u}) + \mu^{(m)} (\vec{\nabla} \vec{u})^T\right)$$

where $\vec{u}$ is the fluid velocity, $\mu^{(m)}$ is the dynamic viscosity of the liquid, $C_L^{(m)}$ stands for the lattice heat capacity of the molten material, $\rho_L^{(m)}$ is the density of the molten material, $T_{melt}$ is its melting temperature, $L_m$ is the latent heat for melting, and $P$ represents the total pressure, including the recoil pressure [19, 44]

$$P_r = 0.54 P_0 \exp\left(L_v \frac{T_L^S - T_b}{R_G T_L^S T_b}\right) \qquad (6)$$

where $P_0$ is the atmospheric pressure (i.e. equal to $10^5$ Pa), $L_v$ is the latent heat of evaporation of the liquid [44], $R_G$ is the universal gas constant (8.314 J/(K mole)), $T_L^s$ corresponds to the surface temperature of the melt and $T_b$ is the boiling temperature of iron (~3100 K [45])). The term in Eq. (5) that contains the delta function has been introduced to provide a smooth transition from the solid-to liquid phase and describe



efficiently the resolidification process. As the movement of the $T_{melt}$ isothermal (inside the volume of the irradiated material) is used to determine the resolidification process, calculations of the liquid-surface interface are performed by an appropriate (i.e. Gaussian) representation of the delta function, $\delta\left(T_L - T_{melt}\right) = \dfrac{1}{\sqrt{2\pi}\Delta} e^{-\left[\dfrac{(T_L - T_{melt})^2}{2\Delta^2}\right]}$, where $\Delta$ is in the range of 10-100 K depending on the temperature gradient [19, 46].

A solid material subjected to ultrashort pulsed laser heating at sufficiently high fluences undergoes a phase transition to a superheated liquid which's temperature reaches $0.90T_{cr}$ ($T_{cr}$ being the thermodynamic critical temperature, $T_{cr}$(Fe) = 8500 K) [47]. According to Kelly and Miotello [47], melted material at and beneath the irradiated surface is unable to boil, as the timescale does not permit heterogeneous nucleation. A subsequent homogeneous nucleation of bubbles leads to a rapid transition of the superheated liquid to a mixture of vapour and liquid droplets that are ejected from the bulk material (a process referred to as phase explosion). This is proposed as a material removal mechanism and it is assumed that phase explosion occurs when the lattice temperature is equal or greater than $0.90T_{cr}$ [19, 47-50]. Hence, to incorporate the mass removal mechanism, it is assumed in our model that all material points that undergo phase explosion are removed.

## 4. Simulations

The parameters that are used in the numerical solution of the governing set of equations (1)-(6) are listed in Table I. A finite-difference method in a staggered grid is employed to numerically solve the heat transfer equations [19], including phase changes (melting, evaporation) [19, 30] and resolidification. The horizontal and vertical velocities are defined in the center of the horizontal and vertical cells faces, where the pressure and temperature fields are defined in the cell center (see description in [19, 35]). Similarly, all temperature dependent quantities (i.e. heat capacity, mass density, etc.) are defined in the cell center. While second-order finite difference schemes appear to be accurate for one laser pulse, where the surface topography profile is not modified substantially, finer



meshes and higher-order methodologies are performed for more complex profiles [51, 52]. Furthermore, techniques that assume moving boundaries (i.e. solid-liquid interface) are employed [53]. The hydrodynamic equations are solved in the sub-region that contains either solid or molten material. To include the "hydrodynamic" effect of the solid domain, material in the solid phase is modelled as an extremely viscous liquid ($\mu_{solid} = 10^5 \mu_{liquid}$), which results in velocity fields that are infinitesimally small.

**Table I.** Simulation parameters chosen for 100Cr6 steel

| Parameter | Value |
|---|---|
| $A$ [s$^{-1}$ K$^{-2}$] | 0.98×10$^7$ (from fitting) |
| $B$ [s$^{-1}$ K$^{-1}$] | 2.8×10$^{11}$ (from fitting) |
| $\kappa_{e0}$ [Wm$^{-1}$K$^{-1}$] | 46.6 [37] |
| $C_L$ [J/kg$^{-1}$K$^{-1}$] | 475 [37] |
| $C_L^{(m)}$ [J/kg$^{-1}$K$^{-1}$] | 748 [44] |
| $T_{melt}$ [K] | 1811 [54] |
| $T_0$ [K] | 300 |
| $T_{cr}$ [K] | 8500 [45] |
| $\rho_L^{(m)}$ [kg/m$^3$] | 6900 [44] |
| $\rho_L^{(s)}$ [kg/m$^3$] | 7700 [37] |
| $\mu^{(m)}$ [Pa s] | 0.016 [55] |
| $\sigma$ [Nm$^{-1}$] | 1.93-1.73×10$^{-4}$($T_L$-$T_{melt}$)K$^{-1}$ [56] |
| $L_v$ [J/g] | 6088[44] |
| $L_m$ [J/g] | 276 [44] |
| $w_0$ [µm] | 15 |
| $\tau_p$ [fs] | 30/170 |

At time $t = 0$, both electron and lattice temperatures are set to room temperature (300 K). Non-slipping conditions (i.e. the spatial velocity field is zero everywhere) are applied on



the solid-liquid interface while the condition $\mu^{(m)} \frac{\partial u_z}{\partial z} = \frac{\partial \sigma}{\partial T_L^S} \frac{\partial T_L^S}{\partial x}$ is applied on the upper and

free surface of the material. Besides the surface tension $\sigma$, the lattice temperature at the

surface, $T_L^S$, is used [19]. Fluence values $F$ in the range 0.1 to 1 J/cm$^2$ are considered in

the simulations. As the material is subjected to irradiation by multiple laser pulses, Eqs.

(1)-(6) are solved in a three dimensional Cartesian coordinate system and the energy

absorption in subsequent irradiation is modelled by considering a ray tracing approach to

compute the absorbed and reflected part in a modified profile (resulting from fluid

dynamics and resolidification).

The variable grid size taken for simulations is 5 nm (vertical dimension) and 10 nm

(horizontal dimension) in the region where material melts while the grid size is [10 nm ×

20 nm] elsewhere. Therefore, the irradiated region is split into two sub-regions to

accommodate solid and molten material. The temporal calculation step is adapted so that

the stability Neumann condition in two dimensions is satisfied [57]. Regarding the

material removal simulation, in each time step lattice and electron temperatures are

computed. If the lattice temperature reaches ~$0.90T_{cr}$ (=7650 K), mass removal is

assumed [50]. In that case, the associated nodes on the mesh are eliminated and new

boundary conditions of the aforementioned form on the new surface are enforced. In

order to preserve the smoothness of the surface that has been removed and to allow an

accurate and non-fluctuating value of the computed curvature and surface tension

pressure, a specific fitting methodology is pursued (see description in [19, 35]).

## 5. Results and discussion

Before the surface modification process due to laser irradiation is analysed, the

electron and lattice temperature evolutions are investigated to compare the thermal

response of bulk 100Cr6 steel and Fe for one laser pulse. The optical properties of

100Cr6 and Fe at various wavelengths are taken from measurements (for 100Cr6 from

spectral ellipsometry) or literature values (for Fe) [41]. Results for $T_e$ and $T_L$ are

illustrated in Fig. 4. As explained above, phase explosion and mass removal are



manifested by using the lattice temperature as a criterion. Hence, the methodology to model the underlying process is through removal of all points with lattice temperatures above $0.90T_{cr}$. In Figs. 4a,b, the evolution of $T_e$ and $T_L$ on the surface ($z = 0$ and $x = y = 0$) is illustrated at all points in time, unless the lattice temperature exceeds the phase

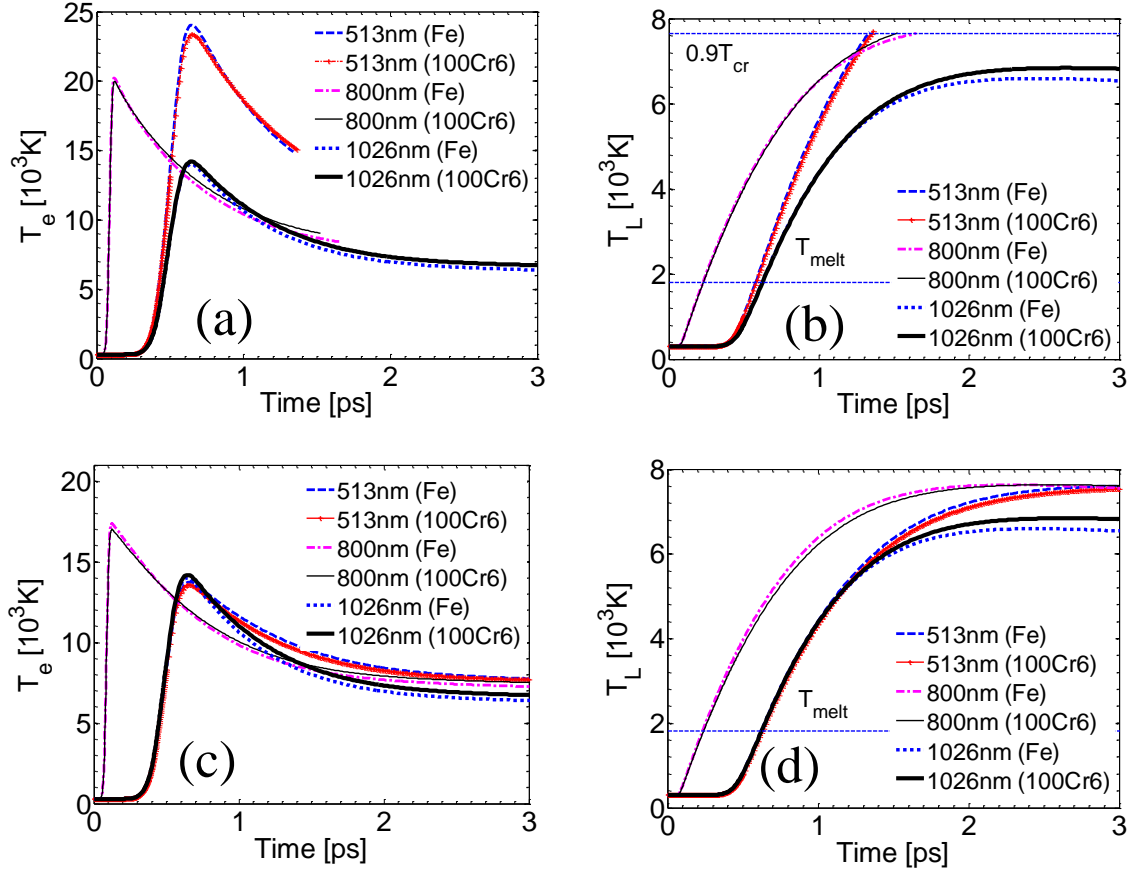

FIG. 4 (Color Online) Evolution of the electron temperature $T_e$ (a) and the lattice temperature $T_L$ (b) for 100Cr6 and Fe at three different laser wavelengths on the surface ($z$=0 and $x$=$y$=0). Evolution of the electron temperature $T_e$ (c) and the lattice temperature $T_L$ (d) for 100Cr6 and Fe on new surface after mass removal ($\lambda_L$ = 513 nm, $\tau_p$ = 170 fs), ($\lambda_L$ = 800 nm, $\tau_p$=30 fs), ($\lambda_L$ = 1026 nm, $\tau_p$ = 170 fs), ($x = y = 0$) ($F = 0.25$ J/cm$^2$) (see enlarged regions in the Supplementary Material).

explosion threshold; in that case, thermal response is not displayed as the lattice point is removed and no longer contributes to the physical processes. The temperature evolution



is displayed at depths at $z = 11.5$ nm and 4.8 nm (for 100Cr6) and $z = 11.8$ nm and 3.5nm (for Fe) for $\lambda_L = 513$ nm and 800 nm, respectively, in Figs. 4c,d. On the other hand, for $\lambda_L = 1026$ nm, $T_e$ and $T_L$ are evaluated at $z = 0$ for both materials. These points correspond to the positions of the material surface following irradiation by one pulse and after mass removal was assumed. It is noted that simulation results have been obtained assuming $\tau_p = 170$ fs (for $\lambda_L = 513$ nm and 1026 nm) and 30 fs (for $\lambda_L = 800$ nm).

Simulations show (Fig. 4a and Fig. S1a in the Supplementary Material) that for $\lambda_L = 513$ nm contrary to $\lambda_L = 1026$ nm, the maximum $T_e$ is always higher for Fe compared to 100Cr6. This is explained in terms of the relative reflectivity and absorption coefficient for Fe and 100Cr6. More specifically, the reflectivity of Fe is larger at $\lambda_L = 513$ nm and 1026 nm (Fig. 3c). Therefore, enhanced laser energy absorption is expected at those values compared to the $T_e$ for 100Cr6. Nevertheless, a larger absorption coefficient for laser irradiation of 100Cr6 (Fig. 3c) leads to a smaller electron energy on the surface [33] that produces larger maximum values of $T_e$ for Fe. This is evident for $\lambda_L = 513$ nm due to the 10% difference of the absorption coefficient (i.e. $\alpha_{Fe} = 7.105 \times 10^5$ cm$^{-1}$ [41], and $\alpha_{100Cr6} = 6.472 \times 10^5$ cm$^{-1}$). In contrast, the small difference in absorption coefficient between the two materials for $\lambda_L = 1026$ nm does not allow a similar trend and therefore the maximum value of the electron temperature is higher for 100Cr6 (Fig. 4c and Fig. S1c in the Supplementary Material). Finally, a smaller reflectivity of Fe (at $\lambda_L = 800$ nm) followed by a larger absorption coefficient of the laser energy for 100Cr6 also produces larger maximum values of $T_e$ for Fe at this wavelength. The difference of the electron-phonon scattering processes and energy exchange for Fe and 100Cr6 is influenced by the bigger heat conductivity of Fe ($\kappa_{e0} = 80.4$ Wm$^{-1}$K$^{-1}$) compared to the 100Cr6 steel ($\kappa_{e0} = 46.6$ Wm$^{-1}$K$^{-1}$ [37]). As a result, the electrons diffuse somewhat faster to deeper parts of the material in Fe that leads to larger maximum $T_L$ upon energy exchange for 100Cr6 compared to the predicted values for Fe (Fig. 4d). Enlarged figures that depict the evolution of $T_e$ and $T_L$ are illustrated for each case in the Supplementary Material (Fig. S1).

The dynamics of the volume of the material that will undergo a phase transition to the liquid phase is described by Navier-Stokes equations as stated in the previous sections. It is evident that the portion of the material in the molten phase is related to both the single



photon energy and the fluence. To estimate the maximum depth of the molten region (at $x = y = 0$)), simulation results after irradiation of 100Cr6 with a laser beam of three different wavelengths [$\lambda_L = 513$ nm ($\tau_p = 170$ fs), $\lambda_L = 800$ nm ($\tau_p = 30$ fs), and $\lambda_L = 1026$ nm ($\tau_p = 170$ fs)] and fluence $F = 0.25$ J/cm$^2$ predict the formation of a molten volume of an average depth equal to approximately 45 nm for $NP = 1$ (see Fig. S2 in the Supplementary Material). A similar approach is followed for $NP > 1$ or other fluence values.

To correlate the excited electron density with possible laser-induced surface structures, the inhomogeneous energy deposition into the irradiated material is computed by the calculation of the product $\eta(\boldsymbol{k}, \boldsymbol{k_i}) \times |b(\boldsymbol{k})|$ as described in the model of J.E. Sipe [17]. In the above expression, $\eta$ represents the efficacy with which the surface roughness at the wave vector $\boldsymbol{k}$ (i.e. normalized wavevector $|\boldsymbol{k}| = \lambda_L/\Lambda$, where $\Lambda$ stands for the predicted structural periodicity) induces inhomogeneous radiation absorption, $\boldsymbol{k_i}$ is the component of the wave vector of the incident laser beam on the material's surface plane and $b$ represents a measure of the amplitude of the surface roughness at $\boldsymbol{k}$. Herein, the periodicity of the rippled structures based on the Sipe model is calculated on basis of fourteen complex valued equations [15] at $\lambda_L = 1026$ nm, 800 nm, and 513 nm, yielding the strongest energy absorption at values of $\Lambda = 1019$ nm, 789 nm, and 506 nm, respectively. More specifically, the calculations were performed for three values of the refractive index ($(n+ik)^2 = \varepsilon = \varepsilon_1 + i\varepsilon_2 = -6.597 + 20.51i$, $-4.328 + 16.17i$, $-3.216 + 10.25i$) corresponding to three wavelengths ($\lambda_L = 1026$ nm, 800 nm, 513 nm), as derived from Fig. 5a.). Standard surface roughness parameters were assumed here for the shape factor ($s = 0.4$) and the filling factor ($f = 0.1$) entered in Sipe's theory [15,17]. The orientation of the structures is *perpendicular* to the polarization vector of the incident beam.



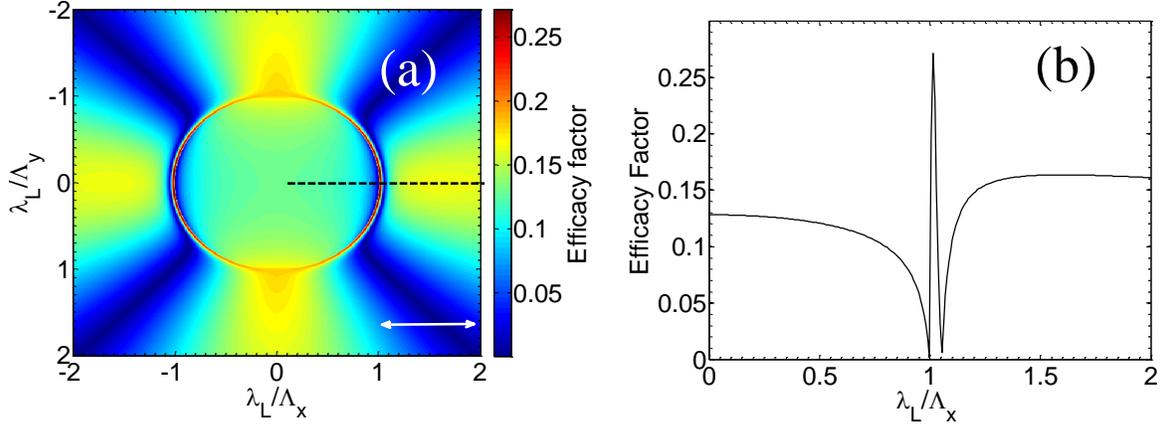

FIG. 5 (Color Online) (a) Efficacy factor ($\eta$) map for refractive index $(n+ik)^2$=(-6.597+20.51$i$) (i.e. $\lambda_L$ = 1026 nm). The double arrow indicates the direction of the polarisation vector of the optical radiation. (b) Efficacy factor $\eta$ along $\lambda_L/\Lambda_x$ [black dashed line in (a)].

Although a multi-pulse feedback mechanism is not included in the original formulation of Sipe's theory, in previous studies, the variation of the rippled periodicity with increasing number of pulses (*NP*) has been simulated by changing the filling factor $f$ (for one pulse, $f$ = 0.1) which represents the fraction of the surface that contains scattering centers [58]. The application of the theory indicates that the increase of $f$ (and correspondingly *NP*) produces a significant increase of $\eta$ along with a small shift of $\Lambda$ to somewhat smaller values. However, there is not a conclusive way to correlate $f$ with *NP* and this constitutes a drawback towards performing a systematic study to correlate the periodicity with *NP*.

In case that the irradiated material is plasmonically active [Re($\varepsilon$) < -1, a condition which is fulfilled for 100Cr6 steel at all three wavelengths, see the values given above], surface plasmon polariton (SPP) excitation may occur and the produced electromagnetic wave can interfere with the incident laser beam, leading to a periodic modulation of the absorbed energy, finally forming ripples at the surface (see [19] and references therein). In fact, SPP excitation and coupling with the incident light is not possible on a flat surface as the laser and SPP dispersion curves do not meet [59]. Therefore, for *NP* = 1, usually rippled structures are not formed on polished surface, unless a defect is already



present at the flat surface during irradiation [21]. Nevertheless, roughness on the surface (generated by laser ablation) is capable of producing sufficient conditions for SPP excitation and laser beam coupling. The desired inhomogeneity on the surface occurs as the lattice temperatures attained in the conditions used in the simulations exceed the boiling temperature $T_b$.

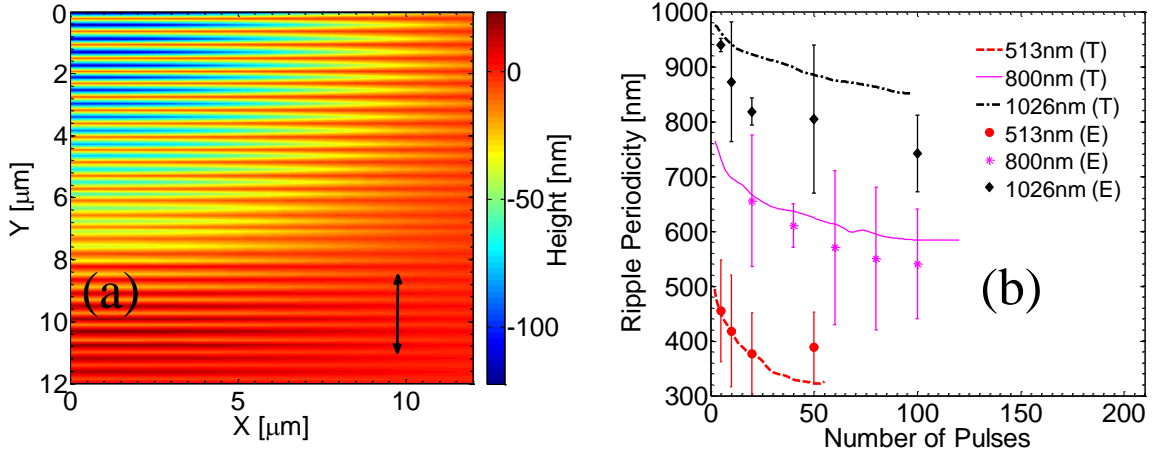

FIG. 6 (Color Online) (a) Rippled surface profile for $\lambda_L = 513$ nm for $NP = 10$ in a 12 μm ×12 μm region. The double arrow indicates the polarization direction of the laser beam. (b) Ripple periodicity for 100Cr6 steel as a function of the number of pulses $NP$ at $\lambda_L$=513 nm, 800 nm, and 1026 nm (simulation results are indicated by '$T$' while '$E$' stands for experimental measurements). The vertical lines indicate the range of ripple periods at the corresponding $NP$ value ($F = 0.25$ J/cm$^2$, $\tau_p = 170$ fs for $\lambda_L = 513$ nm and 1026 nm, and $\tau_p = 30$ fs for $\lambda_L = 800$ nm).

Further irradiation of the corrugated surface (i.e. increase of $NP$) continues to lead to SPP excitation and coupling with the incident beam. For shallow surface modulations, the ripple period can be approximated by the wavelength of the excited surface plasmon polaritons, which was calculated by the simple expression $\lambda_L Re\sqrt{(1+\varepsilon)/\varepsilon}$ [59, 60]. This results in the subwavelength periodicities of 1019 nm, 794 nm, and 506 nm for the three $\lambda_L$. It is evident that there is an excellent agreement here between the predictions of Sipe's model with the simpler SPP model for the ripple (grating) periodicity. Though, the



latter approach is capable of providing a more conclusive picture of the periodicity dependence on *NP*. More specifically, the interaction between the incident beam and the excited SPP wave is estimated assuming modulation of the surface grating structure with increasing *NP*. The solution of the Maxwell's equations along with the requirement of the continuity of the tangential component of the resultant electric field $\vec{E}$ and normal component of $\varepsilon\vec{E}$ on the boundary defined by the grating profile allows determination of the spatial distribution of the electric field and derivation of the dispersion relations [35]. A numerical solution of the equations allows an estimation of the combination of the optimal laser-grating coupling *Λ* and the maximum ripple height. In our model, the coupling of the energy of the incident beam and the SPP is considered in Eq. (1) by a modified source term $S+S_{SP}$ in which $S$ is provided by Eq. (2), while $S_{SP}$ is characterised by a spatially periodic modulation dictated by the grating periodicity [61]. Theoretical results can provide a correlation of *Λ* as a function of the ripple modulation depth that is similar to computations performed for semiconductors [16] or metals [62], which predict for *NP* = 2 periodicity values of 975 nm, 764 nm, and 494 nm for the three $\lambda_L$. The simulations yield the periodicity of the spatially modulated energy that will influence the thermal response of the electron system and, subsequently, the lattice system, eventually leading to melting. Via equations (5), the thermal response of the lattice is transferred directly into the behaviour of the fluid (movement, vorticity) and the induced surface profile upon solidification.

The surface profile for *NP* = 10 is presented in Fig. 6a, which shows clearly the rippled morphology in the irradiated zone. On the other hand, the dependence of the ripple periodicity (close to the spot centre) on *NP* is illustrated for the three different wavelengths in Fig. 6b, where a comparison with experimental observations is also provided. Note that the vertical lines indicate the range of ripple periods at the corresponding *NP* value, which must not be confused with an error bar. Firstly, it is evident that, due to the increased photon energy at lower laser wavelengths, there is a "faster" drop of the periodicity with *NP*. For all wavelengths, this drop is a direct consequence of the modulation depth (height) of the grating-like surface relief, which increases with *NP* and subsequently leads to a decrease of the optimum SPP wavelength [16] as explained in the previous paragraph. The comparison of the theoretical results



with the experimental data demonstrates that they both follow a similar monotonously decreasing trend and reach a plateau. A similar behaviour has also been observed in other materials [19, 21]. To analyse the results, it appears that simulations adequately predict the measured values for 513 nm and 800 nm. More specifically, for small values of *NP*, there is less than 1% (for $\lambda_L = 513$ nm) and 4% (for $\lambda_L = 800$ nm) deviation of the theoretical results from the mean value of the experimental results while the discrepancy for larger values of *NP*, is smaller than 15% (for $\lambda_L = 513$ nm) and 6% (for $\lambda_L = 800$nm). On the contrary, for $\lambda_L = 1026$ nm, for small values of *NP*, the deviation between the theoretical and (mean) experimental value is less than 7% while it is around 15% at larger *NP*. In all aforementioned cases, though, the calculated deviation is smaller as the predicted values lie within the experimental range.

The illustrated results also indicate that theory yields larger periodicities with increasing laser wavelength. This discrepancy is more pronounced for $\lambda_L = 1026$ nm where an abrupt change is followed by a saturation. One possible explanation is related to the oxidation of the material upon repetitive irradiation that changes its dielectric constant [58], the composition of the material and finally its response to successive irradiation. Such a surface modification upon LIPSS formation has recently been studied with nanometer resolution in another metal (Ti), providing evidence for the superficial oxidation during the fs-laser processing in air [63]. This is in line with previous spatially averaged results of Raillard *et al.* for fs-laser irradiated 100Cr6 steel, indicating the formation of iron oxides, chromium oxides, and carbon compounds in the regime of ripple processing [64]. However, for the sake of simplicity and difficulty to model the amount of oxidation, dielectric constant variation due to material nature change was not included in the theoretical framework of this work (a more alternative methodology can be based on the correction of the dielectric constant due to oxidation or by means of the Maxwell–Garnett theory, if nanoparticles are produced on the irradiated material [65]). To show the significance of the oxidation factor in the variation of the dielectric constant, a simple calculation has been performed to highlight the role of the material character change. More specifically, the increase of the refractive index (and increase of the absolute value of the real part of the dielectric constant, $\varepsilon_1^{(\text{non}-\text{oxidised})}$) at higher laser wavelengths for Fe (and 100Cr6) and the corresponding decrease of the refractive index



(and decrease of $\varepsilon_1^{(\text{oxidised})}$) leads to a larger difference $\Delta\varepsilon = \varepsilon_1^{(\text{oxidised})} - \varepsilon_1^{(\text{non-oxidised})}$ [66]. Hence, the expression $\lambda_L Re\sqrt{(1+\varepsilon)/\varepsilon}$ yields a variation $\left(\left(\lambda_{plasmon}^{non-oxidised} - \lambda_{plasmon}^{oxidised}\right)/\lambda_L\right) \sim \Delta\varepsilon$ that confirms that the theoretically computed discrepancy between periodicities (i.e. with and without the consideration of oxidation of the irradiated material) should be more pronounced at higher $\lambda_L$. Therefore, the assumption that the irradiated material continues to behave as a metal with a dielectric constant as described by the results in Fig. 4 yields a rather inaccurate estimation that is more evident at larger laser beam wavelengths. Similar considerations have been explored in other works where an increased real part of the refractive index for propagation of surface plasmons leads to a decrease of the ripple periodicity [58, 67, 68].

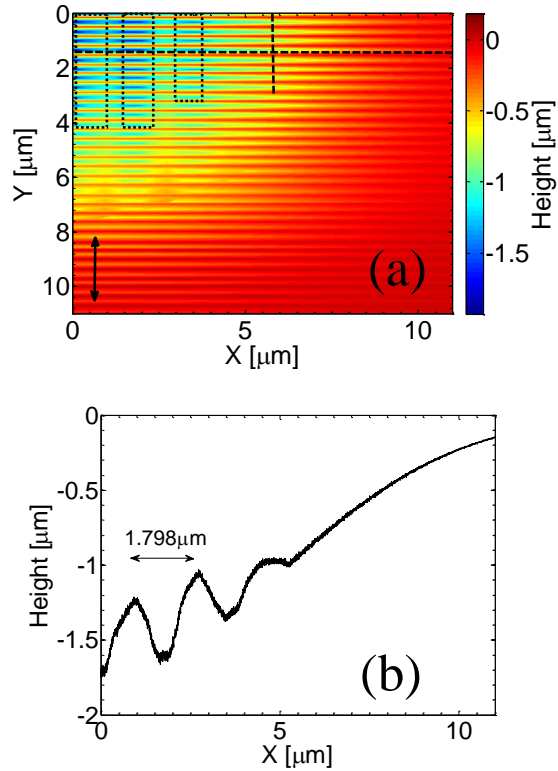



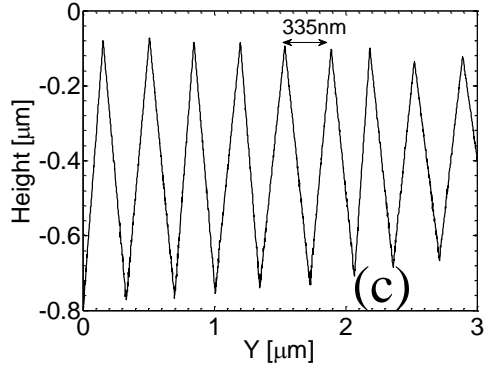

FIG. 7 (Color Online) (a) Surface profile for $\lambda_L$ = 513 nm of 100Cr6 steel for *NP* = 150 (*Dotted* lines indicate grooves). The double arrow indicates the polarization direction of the laser beam. (b) Height profile along *horizontal dashed* line in (a). (c) Height profile along *vertical dashed* line in (a). ($F$ = 0.25 J/cm$^2$, $\tau_p$ = 170 fs).

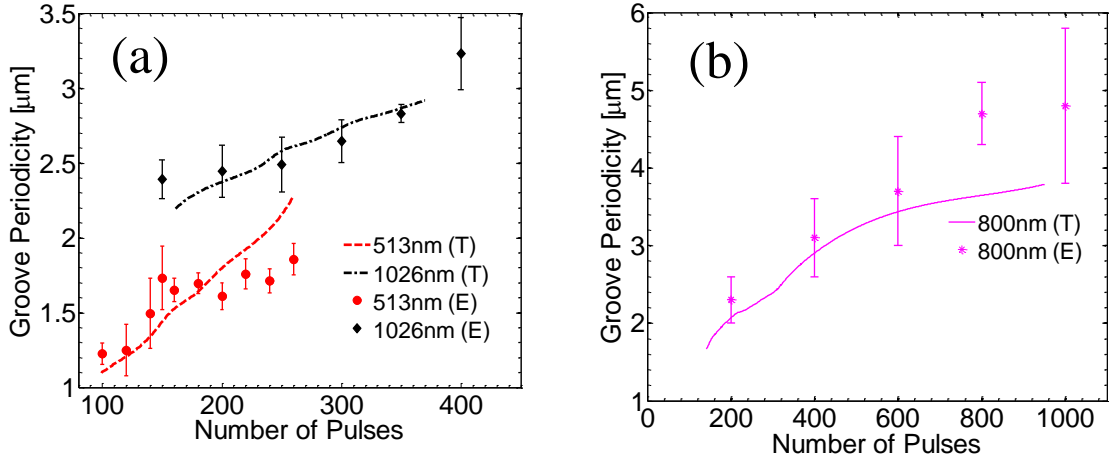

FIG. 8 (Color Online) Groove periodicity for 100Cr6 steel as a function of *NP* at $\lambda_L$ = 513 nm and 1026 nm ($\tau_p$ = 170 fs) (a) and 800 nm ($\tau_p$ = 30fs) (b) (theoretical results are indicated by '*T*' while '*E*' stands for experimental measurements). The vertical lines indicate the range of ripple periods at the corresponding *NP* value ($F$ = 0.25 J/cm$^2$).

As the number of pulses increases, a different type of structure is formed at *NPs* $\geq$ 100. These structures have two characteristics: (i) they have an orientation *parallel* to the polarisation of the electric field of the incident laser beam and (ii) they have a periodicity



larger than the wavelength of $\lambda_L$. It is evident that the decreasing periodicity that is observed for the conventional ripples cannot explain the formation of the so-called grooves. Previous works on semiconductors [30] or dielectrics [20] indicated that the origin of the development of the grooves is predominantly driven by hydrodynamics.

More specifically, while ripple periodicity increases at small *NPs*, at larger *NPs* it reaches a plateau. Further irradiation that leads to a larger surface depth (as a result mainly of ablation) fails to yield a sufficient condition for SPP excitation and interference with the incident beam and the predominant mechanism is fluid transport of the molten material along the walls of the ripples (on a curved region). An approach similar to the one followed in a previous work [30] shows that a solution of the Navier-Stokes equation (last equation in Eqs. (5)) yields a development of counter-rotating convection rolls moving on a curved space. These solutions lead to the propagation of stable hydrothermal waves only for a particular value of the frequency of the produced wave. This value also

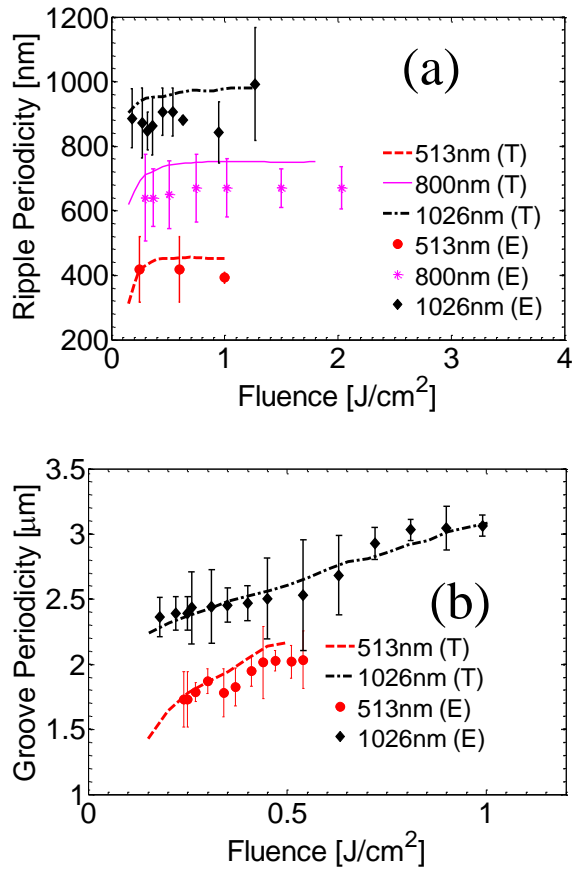



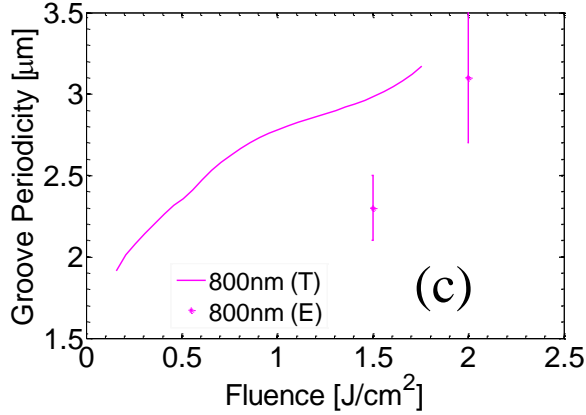

FIG. 9 (Color Online) (a) Ripple periodicity for 100Cr6 chrome steel as a function of fluence ($NP = 10$), (b) Groove periodicity as a function of fluence (at $\lambda_L = 513$nm and 1026 nm and $NP = 150$), (c) Groove periodicity as a function of fluence ($NP = 150$) at $\lambda_L = 800$nm. (Theoretical results are indicated by '$T$' while '$E$' stands for experimental measurements). The vertical lines indicate the range of ripple periods at the corresponding fluence value.

determines the periodicity of the grooves upon solidification while the movement along the walls of the ripples (i.e. a preferred direction) indicates that the orientation of the grooves is perpendicular to the ripples. Simulations predict the formation of grooves between ripples when solidification of the fluid transport of the convection rolls is accomplished. Calculations show that the periodicity of the stable convection rolls is larger than $\lambda_L$ [20, 30]. Certainly, grooves are pronounced in regions where the intensity of the laser beam is higher, therefore, a picture derived from simulations ($NP = 150$) illustrates coexistence of both ripples and grooves (Fig. 7a). While ripples have a sub-wavelength periodicity, grooves are oriented perpendicularly to the ripples with a substantially larger periodicity. Results are illustrated in Fig. 7a for $\lambda_L = 513$ nm (yielding periodicities of 335 nm and 1798 nm for ripples and grooves, respectively). The height profile inside a well (region between two ripples) as a result of the development of convection rolls and formation of grooves is presented in Figs. 7b,c. An investigation of the correlation of the groove periodicity and number of pulses shows that contrary to the monotonous decrease periodicity of the ripples with increasing $NP$, the grooves are



characterized by an increasing periodicity (Fig. 8). This is attributed to enhanced thermal gradients and larger vorticity of the produced hydrothermal waves. To analyse the results in Fig.8, apart from the efficiency of the model to predict the trend, the deviation of the simulated from the measured values for has been quantified for all wavelengths. More specifically, for small values of $NP$, there is less than 11% (for $\lambda_L = 513$ nm) and 7% (for $\lambda_L = 800$ nm) deviation of the theoretical results from the mean value of the experimental results while the discrepancy for larger values of $NP$, is smaller than 18% (for $\lambda_L = 513$ nm and for $\lambda_L = 800$nm); furthermore, the discrepancy is smaller than 3% for all $NP$ for $\lambda_L = 1026$ nm. In all aforementioned cases, though, the calculated deviation is smaller as the predicted values lie within the experimental range.

Theoretical calculations for ripple and groove periodicities illustrate remarkable results of their change with increasing fluence ($F = 0.2$ to 1 J/cm$^2$) at constant number of pulses. The ripple periodicity reaches a saturation value at larger fluences (Fig. 9a) while a continuously rising groove periodicity occurs with increasing fluence (Fig. 9b, for $\lambda_L = 513$ nm and $\lambda_L = 1026$ nm). The latter can be explained by the strength of the hydrothermal waves that constitute the main factor that generate the grooves. These hydrodynamics related effects become stronger at larger fluences which lead to stronger shear stresses on the material and therefore increasingly larger periodicities. It is noted that a similar monotonous trend for the groove dependence on the fluence is predicted for $\lambda_L = 800$ nm (Fig. 9c). Nevertheless, contrary to the agreement of the trend in the groove dependence on fluence with the experimental results for $\lambda_L = 513$ nm and $\lambda_L = 1026$ nm, the small number of experimental data for $\lambda_L = 800$ nm does not allow a convincing conclusion. More specifically, at peak fluences smaller than 1.5 J/cm$^2$, grooves were not observed experimentally for $NP = 150$. Furthermore, experiments at fluences exceeding 2 J/cm$^2$ were not performed in this set of experiments.

A analysis has been performed to quantify the deviation of the theoretical and experimental data and test the efficiency of the model to predict the dependence of the ripple/groove periodicities on various fluence values: For ripples, for all values of the fluence, for $\lambda_L = 513$ nm and $\lambda_L = 800$ nm, there is a less than 15% and 11%, respectively, deviation of the theoretical results from the mean value of the experimental results. On the contrary, for $\lambda_L = 1026$ nm, there is less than 4% (for small values of the fluence) and



13% (for large values of the fluence). On the other hand, for grooves, for $\lambda_L = 513$ nm and $\lambda_L = 1026$ nm, for small fluence values, there is a less than 2% and 1% deviation between the theoretical results and the mean value of the experimental results. Similary, those estimates are 7% and 3% for larger fluence values for the two laser beam wavelengths, respectively. As mentioned in previous paragraphs, the calculated deviations are smaller as the predicted values lie within the experimental range.

The above description of the mechanism that leads to the formation of the aforementioned structures suggests that the proposed theoretical framework can be part of a consistent methodology towards efficient laser processing of steel assuming the large variety of potential applications These investigations not only provide new insights into the mechanism that characterises laser-matter interaction but also offer a systematic methodology towards laser processing of steel surfaces with important applications.

## 6. Conclusions

A detailed theoretical model was presented which is aimed to account for the surface modification and the plasmon-generated-periodic surface structure formation (ripples) and hydrodynamics-driven supra-wavelength structures (grooves) observed upon irradiation of 100Cr6 steel with ultrashort laser pulses. Theoretical simulations provided a thorough picture of the transition from ripples to grooves with increasing number of pulses at different photon energies. The modelling approach and the capability to predict the trend of the size change of the induced periodic structures firstly support a description of the underlying physical processes and secondly allow a systematic methodology of processing 100Cr6 with femtosecond pulsed lasers. Therefore, the model could constitute a part of a more general and accurate theoretical framework that will enable tailoring the morphology of a surface as well as the bulk at the nanoscale according to the demands of various applications.



## Acknowledgement


This work has been supported by the project *LiNaBioFluid*, funded by the European Union's H2020 framework programme for research and innovation under Grant Agreement No. 665337. The authors would like to thank S. Binkowski (BAM 6.3) for polishing the 100Cr6 samples, S. Benemann (BAM 6.1) for SEM, and A. Hertwig (BAM 6.7) for the ellipsometric measurements. G.D.T, A.M, E.Sk. and E.St. also acknowledge financial support from *Nanoscience Foundries and Fine Analysis* (NFFA) - Europe H2020-INFRAIA-2014-2015 (Grant agreement No 654360).


## References


[1]     A. Y. Vorobyev and C. Guo, Laser & Photonics Reviews **7**, 385 (2012).

[2]     V. Zorba, L. Persano, D. Pisignano, A. Athanassiou, E. Stratakis, R. Cingolani, P. Tzanetakis, and C. Fotakis, Nanotechnology **17**, 3234 (2006).

[3]     V. Zorba, E. Stratakis, M. Barberoglou, E. Spanakis, P. Tzanetakis, S. H. Anastasiadis, and C. Fotakis, Advanced Materials **20**, 4049 (2008).

[4]     D. Bäuerle, *Laser processing and chemistry* (Springer, Berlin; New York, 2000), 3rd rev. enlarged edn.

[5]     J.-C. Diels and W. Rudolph, *Ultrashort laser pulse phenomena : fundamentals, techniques, and applications on a femtosecond time scale* (Elsevier / Academic Press, Amsterdam ; Boston, 2006), 2nd edn.

[6]     E. L. Papadopoulou, A. Samara, M. Barberoglou, A. Manousaki, S. N. Pagakis, E. Anastasiadou, C. Fotakis, and E. Stratakis, Tissue Eng Part C-Me **16**, 497 (2010).

[7]     Z. B. Wang, M. H. Hong, Y. F. Lu, D. J. Wu, B. Lan, and T. C. Chong, Journal of Applied Physics **93**, 6375 (2003).

[8]     R. Böhme, S. Pissadakis, D. Ruthe, and K. Zimmer, Applied Physics A-Materials Science & Processing **85**, 75 (2006).

[9]     I. Paradisanos, C. Fotakis, S. H. Anastasiadis, and E. Stratakis, Applied Physics Letters **107**, 111603 (2015).

[10]     J. Bonse, S. Höhm, S. V. Kirner, A. Rosenfeld, and J. Krüger, IEEE J. Sel. Top. Quant. Electron. **23**, 9000615 (2017).

[11]     J. Bonse, R. Koter, M. Hartelt, D. Spaltmann, S. Pentzien, S. Höhm, A. Rosenfeld, and J. Krüger, Applied Physics A-Materials Science & Processing **117**, 103 (2014).

[12]     E. Skoulas, A. Manousaki, C. Fotakis, and E. Stratakis, Scientific Reports **7**, 45114 (2017).

[13]     G. D. Tsibidis, E. Stratakis, and K. E. Aifantis, Journal of Applied Physics **111**, 053502 (2012).

[14]     E. Stratakis, A. Ranella, and C. Fotakis, Biomicrofluidics **5**, 013411 (2011).

[15]     J. Bonse, M. Munz, and H. Sturm, Journal of Applied Physics **97**, 013538 (2005).

[16]     M. Huang, F. L. Zhao, Y. Cheng, N. S. Xu, and Z. Z. Xu, ACS Nano **3**, 4062 (2009).

[17]     J. E. Sipe, J. F. Young, J. S. Preston, and H. M. van Driel, Physical Review B **27**, 1141 (1983).

[18]     Z. Guosheng, P. M. Fauchet, and A. E. Siegman, Physical Review B **26**, 5366 (1982).

[19]     G. D. Tsibidis, M. Barberoglou, P. A. Loukakos, E. Stratakis, and C. Fotakis, Physical Review B **86**, 115316 (2012).

[20]     G. D. Tsibidis, E. Skoulas, A. Papadopoulos, and E. Stratakis, Physical Review B **94**, 081305(R) (2016).

[21]     J. Bonse, A. Rosenfeld, and J. Krüger, Journal of Applied Physics **106**, 104910 (2009).





[22]    T. J. Y. Derrien, T. E. Itina, R. Torres, T. Sarnet, and M. Sentis, Journal of Applied Physics **114**, 083104 (2013).

[23]    M. Barberoglou, G. D. Tsibidis, D. Gray, E. Magoulakis, C. Fotakis, E. Stratakis, and P. A. Loukakos, Applied Physics A: Materials Science and Processing **113**, 273 (2013).

[24]    G. D. Tsibidis, E. Stratakis, P. A. Loukakos, and C. Fotakis, Applied Physics A **114**, 57 (2014).

[25]    O. Varlamova, F. Costache, J. Reif, and M. Bestehorn, Applied Surface Science **252**, 4702 (2006).

[26]    J. Bonse, J. Krüger, S. Höhm, and A. Rosenfeld, Journal of Laser Applications **24**, 042006 (2012).

[27]    S. Yada and M. Terakawa, Optics Express **23**, 5694 (2015).

[28]    E. Rebollar, J. R. V. de Aldana, J. A. Perez-Hernandez, T. A. Ezquerra, P. Moreno, and M. Castillejo, Applied Physics Letters **100**, 041106 (2012).

[29]    J. JJ Nivas, S. He, A. Rubano, A. Vecchione, D. Paparo, L. Marrucci, R. Bruzzese, and S. Amoruso, Scientific Reports **5**, 17929 (2015).

[30]    G. D. Tsibidis, C. Fotakis, and E. Stratakis, Physical Review B **92**, 041405(R) (2015).

[31]    U. Hermens *et al.*, Applied Surface Science **418**, 499 (2017).

[32]    J. M. Liu, Optics Letters **7**, 196 (1982).

[33]    J. Hohlfeld, S. S. Wellershoff, J. Güdde, U. Conrad, V. Jahnke, and E. Matthias, Chemical Physics **251**, 237 (2000).

[34]    Z. Lin, L. V. Zhigilei, and V. Celli, Physical Review B **77**, 075133 (2008).

[35]    G. Tsibidis and E. Stratakis, Journal of Applied Physics **121**, 163106 (2017).

[36]    G. Kresse and J. Hafner, Physical Review B **48**, 13115 (1993).

[37]    Technical    Data    from    Lucefin    Group    (http://www.lucefin.com/wp-content/files_mf/1.3505100cr6.pdf).

[38]    J. Winter, J. Sotrop, S. Borek, H. P. Huber, and J. Minar, Physical Review B **93**, 165119 (2016).

[39]    Y. P. Ren, J. K. Chen, and Y. W. Zhang, Journal of Applied Physics **110**, 113102 (2011).

[40]    A. M. Chen, H. F. Xu, Y. F. Jiang, L. Z. Sui, D. J. Ding, H. Liu, and M. X. Jin, Applied Surface Science **257**, 1678 (2010).

[41]    P. B. Johnson and R. W. Christy, Physical Review B **9**, 5056 (1974).

[42]    A. D. Rakic, A. B. Djurisic, J. M. Elazar, and M. L. Majewski, Applied Optics **37**, 5271 (1998).

[43]    N. W. Ashcroft and N. D. Mermin, *Solid State Physics* (Holt, New York,, 1976).

[44]    V. Semak and A. Matsunawa, Journal of Physics D-Applied Physics **30**, 2541 (1997).

[45]    J. Emsley, *The Elements* (Clarendon Press ;Oxford University Press, Oxford, New York, 1991), 2nd edn.

[46]    N. M. Bulgakova, R. Stoian, A. Rosenfeld, I. V. Hertel, W. Marine, and E. E. B. Campbell, Applied Physics A **81**, 345 (2005).

[47]    R. Kelly and A. Miotello, Applied Surface Science **96-98**, 205 (1996).

[48]    N. M. Bulgakova and I. M. Bourakov, Applied Surface Science **197**, 41 (2002).

[49]    N. M. Bulgakova, A. V. Bulgakov, I. M. Bourakov, and N. A. Bulgakova, Applied Surface Science **197**, 96 (2002).

[50]    J. K. Chen and J. E. Beraun, J Opt A-Pure Appl Op **5**, 168 (2003).

[51]    Y. Morinishi, O. V. Vasilyev, and T. Ogi, Journal of Computational Physics **197**, 686 (2004).

[52]    Y. Morinishi, T. S. Lund, O. V. Vasilyev, and P. Moin, Journal of Computational Physics **143**, 90 (1998).

[53]    M. Zerroukat and C. R. Chatwin, Journal of Computational Physics **112**, 298 (1994).

[54]    R. Le Harzic, D. Breitling, M. Weikert, S. Sommer, C. Föhl, S. Valette, C. Donnet, E. Audouard, and F. Dausinger, Applied Surface Science **249**, 322 (2005).

[55]    M. Korolczuk-Hejnak, High Temperature **52**, 667 (2014).

[56]    J. Brillo and I. Egry, Journal of Materials Science **40**, 2213 (2005).

[57]    H. J. Wang, W. Z. Dai, and L. G. Hewavitharana, International Journal of Thermal Sciences **47**, 7 (2008).

[58]    J. Bonse, S. Höhm, A. Rosenfeld, and J. Krüger, Applied Physics A-Materials Science & Processing **110**, 547 (2013).

[59]    H. Raether, *Surface plasmons on smooth and rough surfaces and on gratings* (Springer-Verlag, Berlin ; New York, 1988), Springer Tracts in Modern Physics, 111.

[60]    T. J. Y. Derrien, J. Krüger, and J. Bonse, Journal of Optics **18**, 115007 (2016).

[61]    J. C. Wang and C. Guo, Journal of Applied Physics **102**, 053522 (2007).

[62]    G. D. Tsibidis, E. Skoulas, and E. Stratakis, Optics Letters **40**, 5172 (2015).





[63]     S. V. Kirner, T. Wirth, H. Sturm, J. Krüger, and J. Bonse, Journal of Applied Physics **122,** 104901 (2017).

[64]     B. Raillard, L. Gouton, E. Ramos-Moore, S. Grandthyll, F. Müller, and F. Müklich, Surface & Coatings Technology **207**, 102 (2012).

[65]     T. Y. Hwang and C. Guo, Journal of Applied Physics **108,** 073523 (2010).

[66]     M. R. Querry, Contractor Report CRDC-CR-85034  (1985).

[67]     T. J.-Y. Derrien, R. Koter, J. Krüger, S. Höhm, A. Rosenfeld, and J. Bonse, Journal of Applied Physics **116**, 074902 (2014).

[68]     A. Y. Vorobyev, V. S. Makin, and C. Guo, Journal of Applied Physics **101**, 034903 (2007).